\documentclass[12pt,a4paper]{JHEP3}

\pdfoutput=1
\usepackage[latin1]{inputenc}
\usepackage[english]{babel}
\usepackage{amsmath,amssymb}
\usepackage[pdftex]{graphicx}
\usepackage{graphicx}
\usepackage{cite}

\newcommand{\beq}{\begin{equation}}
\newcommand{\eeq}{\end{equation}}
\newcommand{\bea}{\begin{eqnarray}}
\newcommand{\eea}{\end{eqnarray}}
\newcommand{\bdm}{\begin{displaymath}}
\newcommand{\edm}{\end{displaymath}}

\def\as{\alpha_s}

\def\m{{\cal M}}
\def\l{{\cal L}}
\def\ord{{\cal O}}

\def\d{\partial}
\def\dy{\Delta y}
\def\yb{\bar{y}}
\def \d{{\rm d} }
\def \f{f^{\rm gap}}

\title{The dijet cross section with a jet veto}
\author{Rosa Mar\'ia Dur\'an Delgado, Jeffrey R. Forshaw, Simone Marzani  and Michael~H.~Seymour  \\School of Physics \& Astronomy, University of Manchester,\\Oxford Road, Manchester, M13 9PL, U.K.}

\preprint{MAN/HEP/2011/10}


\abstract{We study dijet production in proton-proton collisions with a veto on the emission of a third jet in the rapidity region in between the two leading ones. We resum the leading logarithms in the ratio of the transverse momentum of the leading jets and the veto scale and we match this result to leading-order QCD matrix elements. We find that, in order to obtain sensible results, we have to modify the resummation and take into account energy-momentum conservation effects. 
We compare our theoretical predictions for the gap fraction to experimental data measured by the ATLAS collaboration and find good agreement, although our results are affected by large theoretical uncertainties. We then discuss differences and similarities of our calculation to other theoretical approaches.}

\begin{document}
\section{Introduction}

QCD phenomenology and in particular jet physics are playing a central role in the physics programme of the first period of the LHC running 
\cite{2010w,daCosta:2011ni,atlasveto,Chatrchyan:2011wn,2011me,Chatrchyan:2011qt,Khachatryan:2011zj}.
While the predictions of inclusive jet and dijet cross sections are under good perturbative control, the prediction of the associated final states is much more delicate. A crucial role is played by the colour structure of the hard process, which sets the initial conditions both of the parton shower, which describes the perturbative evolution of the event and the hadronization, which describes the transition of those partons into the hadrons that are seen in the final state (for an overview of this physics, see for example \cite{Buckley:2011ms}). While long viewed as a probe of QCD dynamics, it has only more recently become widely realized that a measurement of hard process colour structure could be an important probe of new physics~\cite{kulesza1,kulesza2,sung,coxfp} (although the idea dates back to at least \cite{Dokshitzer:1991he}).

The most direct probe of the colour structure of a hard process is the probability that it does not radiate into some well-defined region of phase space. In this paper we consider the prototypical process, in which a dijet system is measured and the presence of any additional jets in the rapidity interval between them is vetoed. Provided the threshold for reconstructing and vetoing these jets is in the perturbative regime, the cross section for such events can be calculated perturbatively. However, at every order of perturbation theory, logarithms of the ratio of the hard process scale to this jet veto threshold scale arise, and when this ratio is large it is mandatory to sum these logarithms to all orders. When the ratio of scales is not large, these logarithms do not dominate and fixed order perturbative results are more accurate than the resummed results. To provide a prediction that is valid over all values of the ratio, it is necessary to match the resummed and fixed-order calculations, which is one of the aims of this paper. The jet veto ``gap fraction'' has been measured by the Tevatron and HERA collaborations~\cite{D0gaps, cdfgaps,zeus95,zeus06,H102}  and, very recently, for the first time at the LHC \cite{atlasveto}.

In a previous paper~\cite{Forshaw:2009fz} two of us made a first phenomenological study of this observable, comparing the resummation of soft gluons to a standard event generator (\textsc{Herwig}++~\cite{herwig}). We found that neither approach was completely satisfactory. The parton shower simulation does not contain the full colour structure or any account of the mixing between different colour structures, and neglects contributions coming from loop corrections that do not correspond to a non-emission contribution (so-called Coulomb gluon exchanges). These give a sizeable correction at large enough jet transverse momenta and rapidity separation. 
Moreover, higher than expected non-emission probabilities were found, which has more recently \cite{SchoSeym} been explained as a problem with the way \textsc{Herwig}++ implements the colour structure of hard processes involving gluons.
On the other hand, the resummation approach, which is based on the soft gluon approximation, by itself is not sufficient for currently-probed phase space regions. One important effect that it neglects is energy-momentum conservation: within the soft gluon approximation, the additional jets that one is vetoing could be emitted with a suppression only due to matrix elements, whereas in full QCD such emission would be considerably suppressed by the requirement that it carry away enough energy that it be above the jet veto scale.

In this paper we improve the previous resummed predictions for the gap fraction by modifying them to approximately account for energy-conservation effects and by matching them to the leading QCD order calculation. We also include the first tower of non-global logarithms arising from one gluon emission outside of the gap region.

As well as providing an important understanding of the QCD effects that determine emission patterns, a precise calculational framework would enable colour structures to be used to constrain the physics processes leading to new particle production. This has been particularly thoroughly studied for the Higgs boson in association with two jets \cite{Barger:1994zq,Kauer:2000hi,coxfp}, for which the colour structure has been shown to be identical to the dijet processes that we study here \cite{softgluonshiggs}.

In the remainder of this paper, we define more precisely the observable we will be calculating, the gap fraction. We then discuss the all-orders resummation of the associated leading logarithms, and how to match this result with a fixed-order result. Finally, we compare the matched results with the ATLAS data and with other theoretical predictions, before drawing some conclusions.

\section{The gap fraction}
We are interested in dijet production in proton-proton collisions:
$$
h_1(P_1) +h_2 (P_2)\to j(p_3) +j(p_4) +X\,,
$$
where we veto on the emission of a third jet with transverse momentum bigger than $Q_0 $ in the rapidity region between the two jets. In the present study we fix the veto scale at $Q_0=20$~GeV.
$P_{1,2}$ define the incoming hadron momenta and $p_{3,4}$ the outgoing jet momenta.
We define the gap fraction as the ratio of the cross section for this process over the inclusive rate:
\beq \label{gapfracdef1}
f^{\rm gap} = \frac{\d^2 \sigma^{\rm gap}}{\d Q \, \d Y}\Big/\frac{\d^2 \sigma}{\d Q \, \d Y}.
\eeq
In the Born approximation, the final state consists only of the two hard jets, so every event is a gap event and $\f = 1$. Beyond the Born approximation the leading jets are no longer balanced in transverse momentum. We define $Q$ to be the mean of the transverse momenta of the leading jets $Q=(p_{T3}+p_{T4})/{2}$. This choice, in contrast for instance to the transverse momentum of the leading jet, is more stable under the inclusion of radiative corrections\cite{James}. The rapidity separation is defined by $Y = \Delta y - 2 D$, where $\Delta y=|y_3-y_4|$ is the rapidity separation between the centres of the leading jets and $D$ can be freely chosen. In many previous studies, $D$ was set equal to $R$, the jet radius. The ATLAS collaboration instead measure the gap region from the centres of the jets, i.e.~$D=0$ and thus $Y=\Delta y$. 
It is useful to rewrite Eq.~(\ref{gapfracdef1}) using unitarity:
\beq
\frac{\d^2 \sigma^{\rm gap }}{\d Q \, \d Y } +\frac{\d^2 \sigma^{\rm \overline{gap} }}{\d Q \, \d Y } =\frac{\d^2 \sigma}{\d Q \, \d Y },
\eeq
where we have introduced the complement of the gap cross section, which corresponds to requiring at least one jet harder than $Q_0$ in the rapidity region in between the leading ones.  The gap fraction then becomes
\beq \label{gapfracdef2}
f^{\rm gap} = 1-\frac{\d^2 \sigma^{\rm \overline{gap}}}{\d Q \, \d Y}\Big/\frac{\d^2 \sigma}{\d Q \, \d Y}.
\eeq
Our target in this paper is to evaluate this expression at the first order in the strong coupling (LO) and to match it to the resummed calculation. At this accuracy then Eq.~(\ref{gapfracdef2}) contains only tree level contributions; the numerator is the integrated transverse momentum distribution of the third jet over the gap region and the denominator is simply the Born cross section:
\beq \label{gapfracdef3}
f^{\rm gap}_{LO} = 1-\int_{Q_0}^Q \d k_{T} \,\int_{\rm in} \d y \, \d \phi \, \frac{\d^5 \sigma^{\rm \overline{gap}}}{\d k_{T}\,\d y\, \d \phi\,\d Q \, \d Y}\,\Big/\,\frac{\d^2 \sigma^{\rm born}}{\d Q \, \d Y}+ \ord(\as^2),
\eeq
where $k_{T}$, $y$ and $\phi$ are the transverse momentum, the rapidity and the azimuth of the third jet. The notation $\int_{\rm in}$ implies the integral over the gap region in rapidity and azimuth.
We compute the gap fraction for proton-proton collisions at $7$~TeV using \textsc{Nlojet}++~\cite{nlojet} at leading order. The jets are defined using the anti-$k_t$ algorithm~\cite{antikt} with $R=0.6$. We use the \textsc{Cteq}6.6 parton distribution functions (PDFs)~\cite{cteq66} and adopt the same kinematical cuts as the ATLAS collaboration, requiring all jets to have $p_T>20$~GeV and $|y|<4.4$. 

A fixed-order calculation of the  gap fraction is reliable only at small $\Delta y $ and when $Q$ is of the same order as $Q_0$. As soon as we move away from this region, the leading-order gap fraction decreases rapidly and eventually becomes negative. This unphysical behaviour indicates that the fixed order calculation by itself is not reliable. Large logarithms of the ratio $Q/Q_0$ contaminate the perturbative expansion and they must be resummed to all orders, as discussed in the next section. Also terms proportional to $\Delta y$ (formally equivalent to a logarithm) can be resummed, for instance as in the High Energy Jets (HEJ) framework~\cite{HEJ}. In the limit of large $\Delta y$ and $Q/Q_0$, the cross section is dominated by the singlet exchange component and there is overlap between the logarithms resummed by the approach we describe here and those resummed by the BFKL equation \cite{Forshaw:2005sx}.

\section{The resummed calculation}

The technique for resumming logarithms of the ratio $Q/Q_0$ for the gaps-between-jets cross section has been explained in detail in~\cite{KOS,OS,Oderda,SLL1,SLLind,Forshaw:2009fz}. It relies on the ability to map the real part of loop corrections into a form that is exactly equal and opposite to the phase space integral for real emission. For an observable in which emission is suppressed equally in all angular regions of the event, a global observable, there is an exact cancellation between the real and virtual contributions, such that the result (up to a phase term) is a virtual integral over the part of phase space in which real emission is vetoed.

In the observable we are studying, however, radiation is only suppressed in part of the phase space region and not globally. For this reason ``in-gap'' virtual corrections are not enough to capture even the leading logarithmic accuracy. Radiation outside the gap is prevented from re-emitting back into the gap by the veto requirement, inducing a further real--virtual miscancellation and additional towers of leading logarithms, called non-global logarithms~\cite{nonglobal}. Currently, these contributions can be resummed only in the large $N_c$ approximation\cite{Dasgupta:2002bw,BMS}.
Here, instead we adopt the approach suggested by \cite{SLL1}: we keep the full colour structure but we expand in the number of gluons, real or virtual, outside the gap. It was argued in~\cite{Forshaw:2009fz} that this may be a reasonably convergent expansion, so that the full result is approximated by the contributions arising from only zero or one gluons outside the gap:
\beq
\frac{\d^2 \sigma^{\rm gap }_{\rm res }}{\d Q \, \d Y } =\frac{\d^2 \sigma^{(0)}}{\d Q \, \d Y } +\frac{\d^2 \sigma^{(1)}}{\d Q \, \d Y }+\dots
\eeq
The first contribution to this expansion corresponds to the exponentiation of the one-loop virtual corrections (with no gluon outside the gap).

We define the resummed gap fraction as 
\beq \label{gapres}
f^{\rm gap}_{\rm res} = \frac{\d^2 \sigma^{\rm gap}_{\rm res }}{\d Q \, \d Y}\Big/\frac{\d^2 \sigma^{\rm born}}{\d Q \, \d Y}.
\eeq
Because we are working in the eikonal approximation, additional radiation does not change the Born kinematics and the resummed cross section factorizes into products of resummed partonic contributions and parton luminosity functions:
\beq \label{gapsxsec}
\frac{\d^2 \sigma^{(i)}}{\d Q \, \d Y} = \frac{\rho \pi \as^2}{2 Q S}  \sum_{a,b,c,d}  \frac{1}{1+\delta_{ab}}\frac{1}{1+\delta_{cd}}|\m^{(i)}_{abcd}|^2\mathcal{L}_{ab}(\dy,Q) \Big |_{\Delta y= Y} 
\eeq
with
\beq  \label{difflum}
  \mathcal{L}_{ab}(\dy,Q) =\frac{1}{2 z} \int_{-\yb^+}^{\yb^+} \d \yb \, 
 f_a(\sqrt{z} e^{\yb/2},Q) f_b(\sqrt{z} e^{-\yb/2},Q) \,,
\eeq
where $z=x_1 x_2$, where $x_{1,2}$ are the longitudinal momentum fractions of the incoming partons, $\yb = y_3+y_4$ and $\rho= \frac{4 Q^2}{S}$. The integration limits are
\beq
\yb^+ = {\rm min} \left(\ln \frac{1}{z},  \yb_{\rm cut}\right),
\eeq
where  the value $\yb_{\rm cut}$  is obtained by requiring that both jets are within the calorimeter acceptance.

The resummation of global logarithms is achieved by considering the original four-parton matrix elements dressed by in-gap virtual gluons, with transverse momenta above $Q_0$ and no out-of-gap (real or virtual) gluons. The resummed partonic cross section then has the form
\beq \label{resummedpartonicxsec2}
|\m^{(0)}|^2   =  {\rm tr} \left( He^{- \xi(Q_0,Q) \mathbf{\Gamma}^{\dagger}}Se^{- \xi(Q_0,Q) \mathbf{\Gamma}}\right),
\eeq
where $\xi$ is computed by considering the strong coupling at one loop:
\beq \label{rc}
\xi(k_1,k_2)=\frac{2}{\pi} \int_{k_1}^{k_2} \frac{\d k_T}{k_T} \as(k_T)= \frac{1}{\pi \beta_0} \ln \frac{1+ \as(Q) \beta_0 \ln \frac{k_2^2}{Q^2} }{1+ \as(Q) \beta_0 \ln \frac{k_1^2}{Q^2}},
\eeq
with $\beta_0= \frac{11 C_A- 2 n_f}{12 \pi }$. The matrix $H$ in Eq.~(\ref{resummedpartonicxsec2}) gives the matrix elements of the hard process in some colour  basis, while $S$ is the metric tensor in that colour basis. In an orthonormal basis, as we use throughout this paper, $S=1$.

The soft anomalous dimension is a matrix, with elements $\Gamma_{ij}=\langle e_i|\boldsymbol{\Gamma}|e_j\rangle$ in a basis $\{e_i\}$:
\bea \label{gamma_numeric}
\boldsymbol{\Gamma} &=& -\frac{1}{2}\int_{\rm in} \frac{\d \phi}{2 \pi }\d y\Big[  
\mathbf{t}_a \cdot \mathbf{t}_b\, \omega_{12}+
\mathbf{t}_a \cdot \mathbf{t}_c\, \omega_{13}+
\mathbf{t}_a \cdot \mathbf{t}_d\,\omega_{14} \nonumber \\ &&+
\mathbf{t}_b \cdot \mathbf{t}_c \,\omega_{23}+
\mathbf{t}_b \cdot \mathbf{t}_d \,\omega_{24}+
\mathbf{t}_c \cdot \mathbf{t}_d \,\omega_{34} \Big]+ i \pi \mathbf{t}_a \cdot \mathbf{t}_b\, , \nonumber \\  \omega_{ij}&=&\frac{1}{2}k_T^2\frac{p_i \cdot p_j}{p_i \cdot k \; p_j \cdot k},
\eea
where $\mathbf{t}_i$ is the colour charge of parton $i$\footnote{Note that the mismatch between the indices $1,2,\ldots$ and $a,b,\ldots$ is related to the fact that one must sum over two orientations of the event, e.g.~$i(p_1,\mathbf{t}_a)+j(p_2,\mathbf{t}_b)\to k(p_3,\mathbf{t}_c)+l(p_4,\mathbf{t}_d)$ and $i(p_1,\mathbf{t}_a)+j(p_2,\mathbf{t}_b)\to l(p_3,\mathbf{t}_d)+k(p_4,\mathbf{t}_c)$, as explained in more detail in \cite{Forshaw:2009fz}.}. The soft gluon momentum is labelled $k$. The integrals over the gluon's azimuth and rapidity inside the gap region admit simple analytical expressions if one considers azimuthally symmetric gaps. However, in the current analysis, we are defining the gap from the centres of the leading jets. Thus, the gap region is not just a rectangle in the $(\eta,\phi)$ plane and we have to integrate around the two semi-circular boundaries of the leading jets. Analytical expressions can still be obtained as a power series in $R$~\cite{Appleby:2003sj}. Here instead we decide to keep the full $R$ dependence and perform the integrals numerically when we cannot find  simple analytical results.
For the explicit expressions of the hard scattering matrices $H$ in the various partonic channels we refer to~\cite{Forshaw:2009fz}.

\subsection{Non-global contribution}
We want to estimate the impact of non-global logarithms on the gap fraction. In particular we aim to resum the non-global logarithms that arise as a result of allowing one soft gluon outside the rapidity gap. The general framework in which this calculation is performed is described in~\cite{SLL1,SLLind}, where the case of an azimuthally symmetric gap was considered. As in the global case this led to relatively simple analytical expressions. To include the gap definition used in the ATLAS analysis instead, one has to resort to evaluating most of the integrals numerically. This considerably slows down the calculation, but has a very small effect on the final results. Therefore, for the current work, we decide to include non-global effects as a $K$-factor:
\beq\label{ng_K}
K(Q, \Delta y)= \frac{\frac{\d^2 \sigma^{(0)}}{\d Q \, \d Y } +\frac{\d^2 \sigma^{(1)}}{\d Q \, \d Y }}{\frac{\d^2 \sigma^{(0)}}{\d Q \, \d Y } },
\eeq
where in calculating this ratio we compute the resummed cross sections for azimuthally symmetric gaps, and then use it to multiply the resummed result for zero gluons outside the gap including the exact gap definition. The error this approximation induces is much smaller than the overall uncertainty in the resummed approach, which we estimate below. This approximation does not affect our matching procedure because we are only performing LO matching and non-global logarithms start at $\ord\left(\as^2 \right)$ in the expansion of the gap fraction. We do not need to include effects related to parton recombination due to the particular choice of the jet-algorithm~\cite{Banfi:2005gj,Delenda:2006nf} since we employ the anti-$k_t$ jet algorithm \cite{antikt}.

The calculation of the contribution from one gluon outside the gap is essentially that presented in~\cite{SLL1,SLLind}, except that the final integral over that gluon's momentum is explicitly performed numerically: We briefly recap the results. The soft anomalous dimension for four-parton evolution in the case of azimuthally symmetric gaps reduces to
\beq \label{gammaoperator}
\mathbf{\Gamma} = \frac{1}{2}Y \mathbf{t}_t^2+ i \pi \mathbf{t}_a\cdot \mathbf{t}_b +\frac{1}{4}\rho(Y; |\dy|)(\mathbf{t}_c^2+\mathbf{t}_d^2),
\eeq
where
\beq \label{jetfunc}
\rho(Y;\dy)= \ln \frac{\sinh\left(\dy/2+ Y/2 \right)}{\sinh\left(\dy/2- Y/2 \right)}-Y,
\eeq
and $\mathbf{t}_t=\mathbf{t}_a+\mathbf{t}_c$ is the colour charge matrix corresponding to emission from the total colour exchanged in the $t$ channel.
To obtain the contribution from one gluon outside the gap we must now consider both real and virtual corrections to the four-parton scattering, each dressed with any number of soft gluons:
\beq \label{master1}
|\m^{(1)}|^2 = - \frac{2 }{\pi} \int_{Q_0}^{Q} \frac{\d k_T}{k_T}\as(k_T) \int_{\rm out} \d y \, \left( \Omega_R + \Omega_V\right)\,,
\eeq
where the integrals are over the transverse momentum and rapidity of the real or virtual out-of-gap gluon. The operator to insert this gluon off the external legs is
\beq 
\boldsymbol{D}^{\mu} =  \mathbf{t}_a h_1^{\mu}+\mathbf{t}_b h_2^{\mu}+\mathbf{t}_c h_3^{\mu}+\mathbf{t}_d h_4^{\mu},\quad h_i^{\mu}=\frac{1}{2}k_T\frac{p_i^{\mu}}{p_i \cdot k} , 
\eeq
for real emission, and
\bea
\boldsymbol{\gamma} &=& -\frac{1}{2}\Big[  
\mathbf{t}_a \cdot \mathbf{t}_b\, \omega_{12}+
\mathbf{t}_a \cdot \mathbf{t}_c\, \omega_{13}+
\mathbf{t}_a \cdot \mathbf{t}_d\,\omega_{14}+
\mathbf{t}_b \cdot \mathbf{t}_c \,\omega_{23}+
\mathbf{t}_b \cdot \mathbf{t}_d \,\omega_{24}+
\mathbf{t}_c \cdot \mathbf{t}_d \,\omega_{34} \Big]
, \nonumber \\ \omega_{ij}&=&\frac{1}{2}k_T^2\frac{p_i \cdot p_j}{p_i \cdot k \; p_j \cdot k},
\eea
for virtual emission.
In the case of the out-of-gap gluon being virtual, the subsequent evolution is unchanged from that of the original four-parton system, given by $\mathbf{\Gamma}$ in Eq.~(\ref{gammaoperator}). In the case of real emission we have to consider the colour evolution of a five-parton system~\cite{Kyrieleis:2005dt,5partons}. If we assume the gluon to be emitted on the same side of the event as partons $a$ and $c$, the anomalous dimension is given by
\bea
\mathbf{\Lambda} &=& \frac{1}{2} Y \mathbf{T}^2_t + i \pi \mathbf{T}_a\cdot \mathbf{T}_b+\frac{1}{4} \rho(Y;\Delta y)(\mathbf{T}^2_c+\mathbf{T}^2_d)+\frac{1}{4} \rho(Y; 2|y|)\mathbf{T}^2_k
 \nonumber \\
&+&
 \frac{1}{2}\lambda(Y;|\Delta y|,|y|,\phi)\mathbf{T}_c\cdot \mathbf{T}_k,
\phantom{(99)}\\
\mathbf{T}^2_t&=&(\mathbf{T}_b+\mathbf{T}_d)^2
\eea
and we have introduced the kinematic function
\beq
\lambda(Y;\Delta y,y,\phi) =\frac{1}{2}\ln\frac{\cosh(\dy/2+ y+Y)-\mathrm{sgn}(y) \cos \phi}{\cosh(\dy/2+ y-Y)-\mathrm{sgn}(y) \cos \phi}-Y.
\eeq
The real and virtual out-of-gap emissions, dressed to all orders with in-gap virtual corrections are thus
\bea \label{omega}
 \Omega_R&=& {\rm tr} \left[H e^{- \xi(k_T,Q) \mathbf{\Gamma}^{\dagger}}
 {\mathbf{D}^\mu}^{\dagger}  e^{- \xi(Q_0,k_T) \mathbf{\Lambda}^{\dagger}} e^{- \xi(Q_0,k_T) \mathbf{\Lambda} }\mathbf{D}_\mu
 e^{- \xi(k_T,Q) \mathbf{\Gamma}}\right], \nonumber \\
 \Omega_V &=& {\rm tr} \left[H e^{- \xi(Q_0,Q) \mathbf{\Gamma}^{\dagger}}e^{- \xi(Q_0,k_T) \mathbf{\Gamma}}\boldsymbol{\gamma}e^{- \xi(k_T,Q) \mathbf{\Gamma}}+ {\rm c.c.}\right].
\eea

\begin{figure} 
\begin{center}
\includegraphics[width=0.495\textwidth, clip]{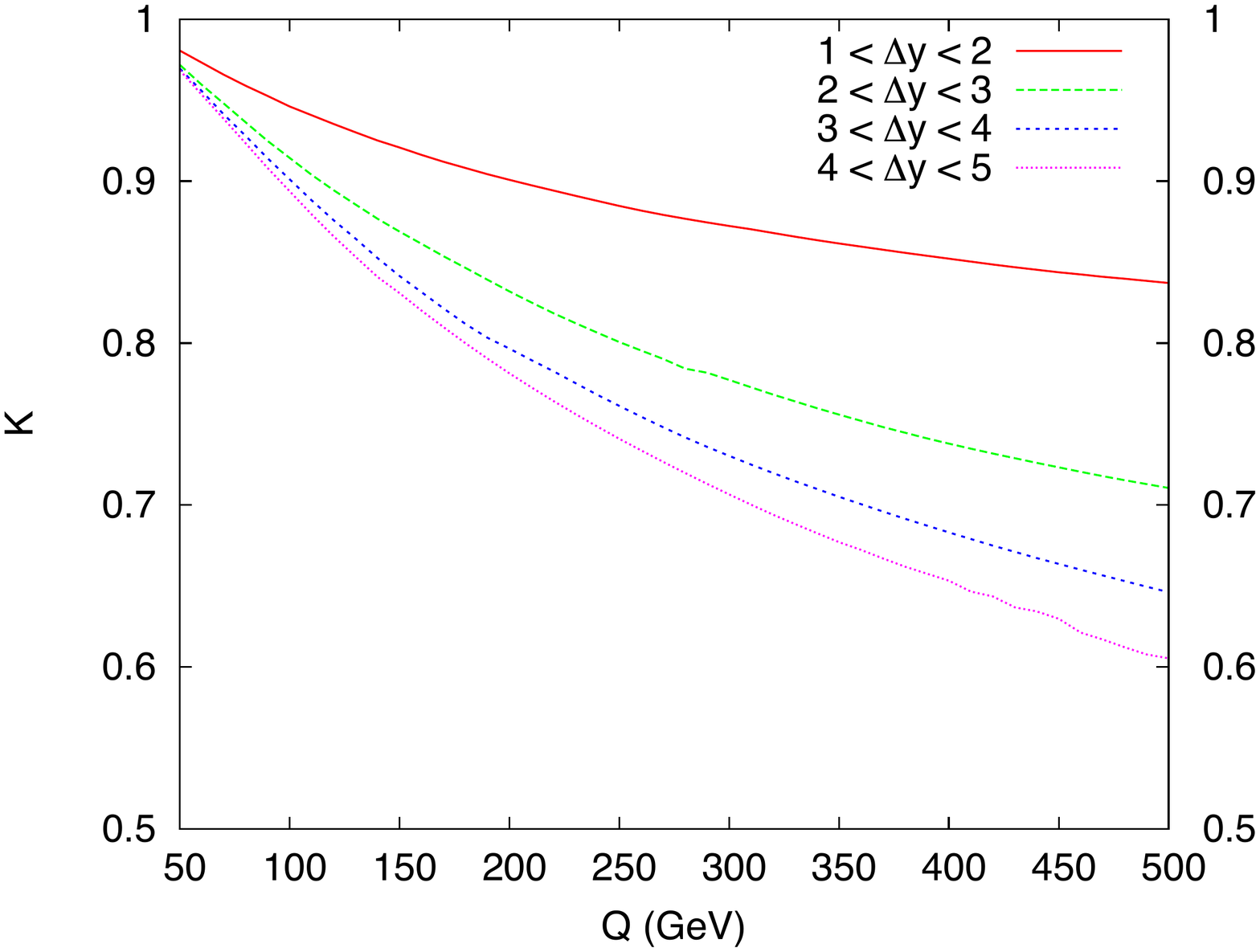}
\includegraphics[width=0.495\textwidth, clip]{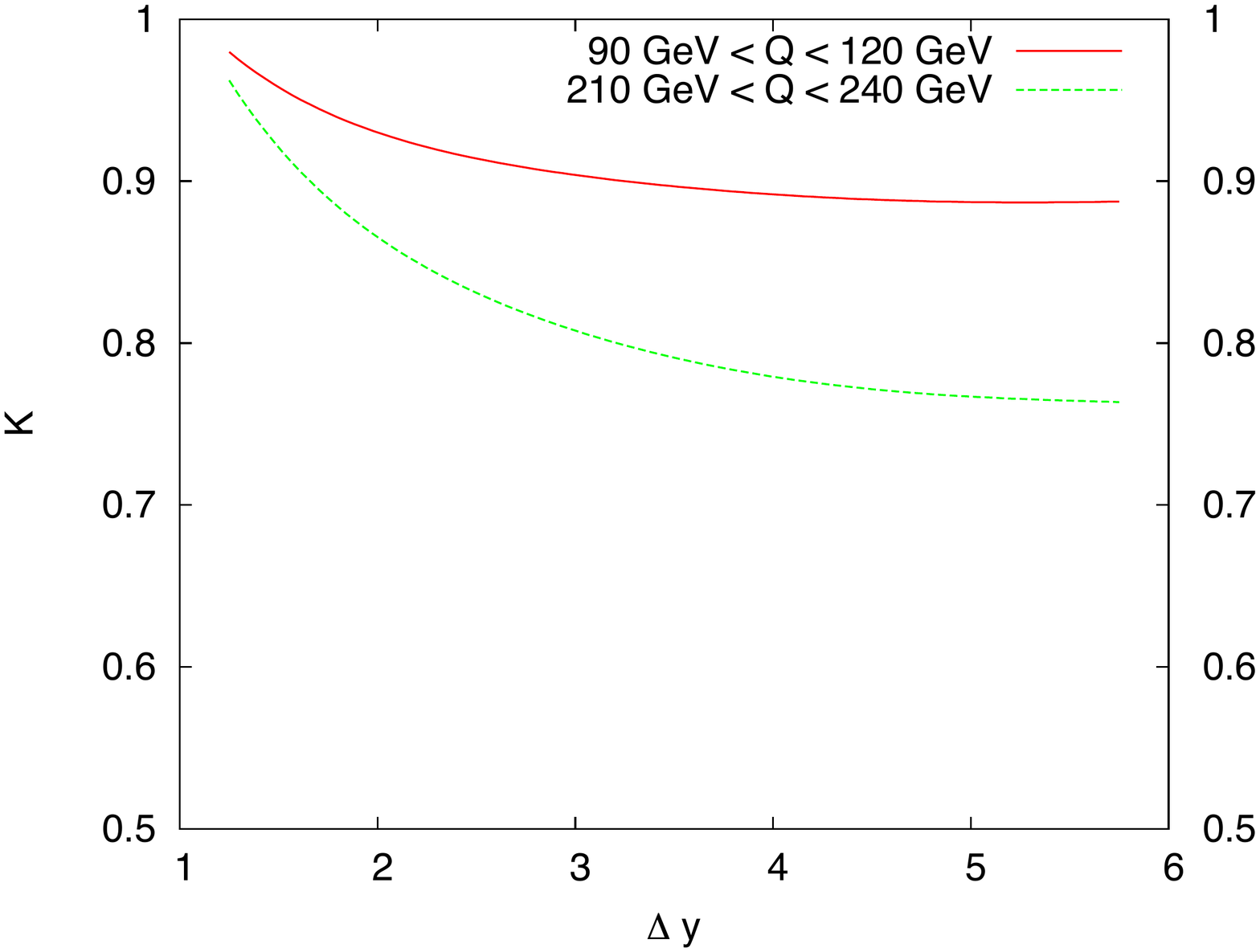}
\caption{The $K$-factor we use to estimated non-global effects, as a function of $Q$ for different values of $\Delta y$ (on the left) and as a function of $\Delta y$, for different values of $Q$ (on the right).}\label{fig:NG}
\end{center}
\end{figure}

The $K$-factor defined in Eq.~(\ref{ng_K}) is plotted in Fig.~\ref{fig:NG} as a function of $Q$ for different rapidity values and as a function of $\Delta y$ for different values of $Q$. In the case of the $Q$ distribution,  we see that the effect is modest for the first rapidity bin, but is typically of the order of 30\% in much of the $Q$ range we study.  Non-global effects are much smaller in the case of the $\Delta y$ distributions for the regions of $Q$ we are considering.

It has been shown~\cite{SLL1,SLLind} that na\"ive QCD coherence is violated at sufficiently high perturbative orders because the Coulomb gluon exchange terms included in $\boldsymbol{\Gamma}$ and $\boldsymbol{\Lambda}$ induce a mis-cancelation between the real and virtual contributions in Eq.~(\ref{master1}). The $y\to\infty$ region therefore gives a finite contribution and, as a consequence, super-leading logarithms ($\as^n\log^{n+1}(Q/Q_0)$) arise at $\ord(\as^4)$ and beyond. The numerical impact of these contributions has been studied in~\cite{Forshaw:2009fz} and found to be generally modest.

\section{Matching}
The resummed calculations are based on the eikonal approximation, in which energy-momentum is not conserved. There is no recoil of the hard lines against the emissions and no account of their effect on the incoming parton momentum fractions at which the parton distribution functions are evaluated. Matching to the full $2\to3$ matrix elements takes into account energy-momentum conservation, at least for the first (hardest) emission. Its energy is taken into account and it is hence is less likely, so the matched gap fraction will be bigger than the one given in Eq.~(\ref{gapres}).

\begin{figure} 
\begin{center}
\includegraphics[width=0.7\textwidth, clip]{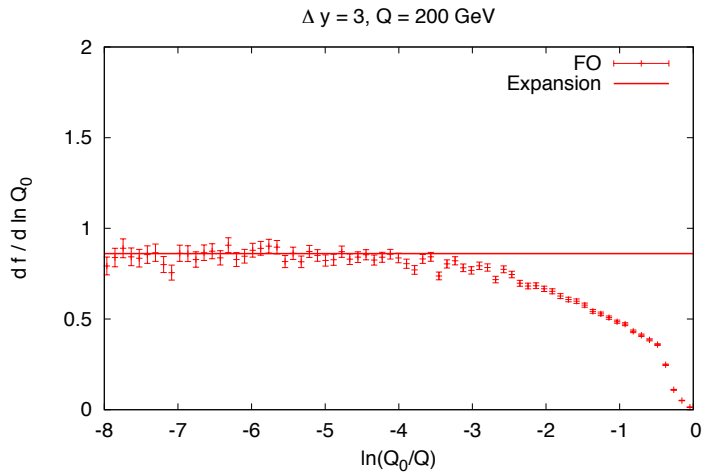}
\caption{The logarithmic derivative of the gap fraction $\frac{\d f^{\rm gap }}{\d \ln Q_0} $ as a function of $\ln Q_0/Q$ for fixed $Q=200$~GeV and $\Delta y =3$ ($\sqrt{S}=7$~TeV).  The solid line is the coefficient obtained by expanding Eq.~(\ref{gapres}) at $\ord (\as)$.}\label{fig:distr}
\end{center}
\end{figure}

In this section we discuss the matching of the resummed gap fraction to the LO calculation. At LO we can write Eq.~(\ref{gapfracdef3}) as follows:
\beq
f^{\rm gap}_{LO} = 1- \frac{2 \as(Q)}{\pi} \left[a_0(Q,Y) \ln \frac{Q}{Q_0}- b_0(Q,Q_0,Y) \right] + \ord( \as^2),
\eeq
where the contribution $b_0$ is now free of large logarithms of $Q/Q_0$.
We want to combine this expression with Eq.~(\ref{gapres}), subtracting the double counted term $1- \frac{2 \as}{\pi}a_0 \ln (Q/Q_0)$. Firstly, we have to verify that the two calculations agree in the asymptotic limit $ \ln (Q/Q_0) \gg 1$. This can be easily achieved considering the logarithmic derivative of the gap fraction:
\beq \label{Fconst}
 \frac{2 \as(Q)}{\pi}a_0(Q,Y) =   \lim_{Q_0 \to 0}\frac{\d}{\d \ln Q_0} f^{\rm gap}_{LO}=-\lim_{Q_0 \to 0}\frac{\d^3 \sigma^{\rm \overline{gap}}}{\d \ln Q_0\,\d Q \, \d Y}\Big/\frac{\d^2 \sigma^{\rm born}}{\d Q \, \d Y}.
\eeq
The result is shown in Fig.~\ref{fig:distr}: the logarithmic derivative of the gap fraction is plotted as a function of $\ln(Q_0/Q)$ for fixed kinematics. In this particular example we have $\Delta y = 3$ and $Q=200$~GeV.  The plot shows that the logarithmic derivative of the gap fraction tends to a constant for large, and negative, values of the logarithm. The numerical value is in agreement with the one obtained by expanding the resummation at $\ord(\as(Q))$:
\bea \label{eikonalexp}
f^{\rm gap}_{\rm res}&=&1- a_0(Q,Y) \xi - a_1(Q,Y) \xi^2 + \dots \nonumber\\
&=& 1- \frac{2 \as(Q) }{\pi} a_0(Q,Y) \left[\ln \frac{Q}{Q_0}+\sum_{n=1}^{\infty}\beta_0^n \, \as(Q)^n \int_{Q_0}^{Q} \frac{\d k_T}{k_T} \ln^n\frac{Q^2}{k_T^2}\right] + \ord(\xi^2)\,. \nonumber \\ 
\eea

\begin{figure} 
\begin{center}
\includegraphics[width=0.7\textwidth,clip]{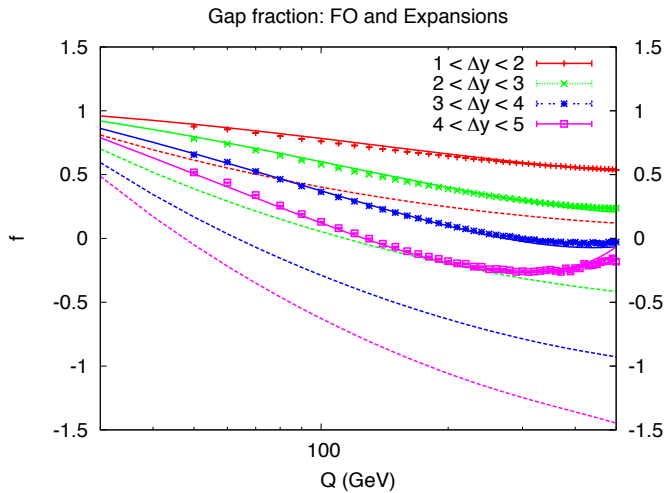}
\caption{The gap fraction at $\ord(\as)$ as a function of $Q$ in different $\Delta y$ bins.
The points are the exact FO calculations, the dashed ones the expansions of the eikonal resummation (Eq.~(\ref{eikonalexp})) and the solid curves correspond to the $\ord(\alpha_s)$ expansion of the modified resummation.}\label{fig:mod}
\end{center}
\end{figure}

The plot in Fig.~\ref{fig:distr} shows that we have control of the logarithms at $\ord(\as)$. However, plotting instead the $Q$ dependence at fixed $Q_0$ for various $\Delta y$ bins, as in Fig.~\ref{fig:mod}, the picture is not so clear. Because the plot is on a logarithmic $x$ axis, we might na\"ively expect the FO result (data points) to asymptotically tend to a straight line, with the same slope as the expansion of the resummation (dashed curve). However, changing the $Q$ values, one changes the momentum fractions and factorization scales of the parton distribution functions and hence the mix of different flavour processes, so one could expect some curvature, but this effect should also be included in the expansion of the resummed results, where some curvature is also seen, so the differences in slope between the data points and dashed curves is really significant. Because the FO curve and the expansion of the eikonal resummation differ so much, a simple matching procedure in which we add together the FO and the resummation and subtract their common term, is bound to fail. It is clear that this issue must be investigated in more detail. The strengthening of the curvature at the highest $Q$ and $\Delta y$ values indicates that we are becoming sensitive to the kinematic limit and we therefore examine the issue of energy-momentum conservation.

\subsection{Energy-momentum conservation}
The resummed cross section (Eq.~(\ref{gapsxsec})) has been obtained in the eikonal limit, i.e.~emitted gluons are considered soft and they do not change the Born kinematics. Even if we are guaranteed that this assumption is sufficient to capture the leading logarithmic behaviour, we are losing important physical effects related to energy-momentum conservation. In particular, because of the choice $Q_0=20$~GeV we are sensitive to emissions of gluons with non-negligible transverse momentum with respect to $Q$. Furthermore, the emission of a gluon requires a finite amount of energy and, for given $Q$ and $\Delta y$, this means we are probing the parton distribution functions at larger values of $x_{1,2}$. Because the PDFs are steeply falling functions of $x$ at large $x$ this can give a considerable suppression well before we reach the edge of phase space.

We would therefore like to go beyond the soft approximation and modify our resummation so that we can capture the correct kinematic behaviour, at least for the hardest (i.e. highest $k_T$) gluon emission. 
In order to do that we study the full kinematics of a $2\to3$ process and using energy-momentum conservation we determine the values of $x_{1,2}$:
\beq \label{modx12}
x_{1,2} = A_{\pm} e^{\pm \yb} \,, \quad {\rm with} \quad
A_{\pm} = \frac{2}{\sqrt{S}}\left[Q \cosh\frac{\Delta y}{2} \pm \bar{Q} \sinh\frac{\Delta y}{2}  + \frac{k_T}{2} e^{\pm  y'}\right]\,.
\eeq
The variables $Q$, $\Delta y$ and $\yb$ are the same as the ones previously defined. We see that $x_{1,2}$ depend on the transverse momentum of the emitted gluon $k_T$ and its rapidity in the partonic centre of mass frame, $y'$. We have also introduced
\beq \label{Qbar}
 \bar{Q}=p_{T3}-p_{T4}= -\frac{k_T}{2}\frac{k_T+ 2 Q \cos \phi}{2 Q + k_T \cos \phi}\,.
 \eeq
 As a consequence the parton luminosities become
\beq  \label{difflum_mod}
 \tilde{\mathcal{L}}_{ab}(\dy,Q,k) =\frac{1}{2 A_+ A_-} \int_{\yb^-}^{\yb^+} \d \yb \, 
 f_a(A_+ e^{\yb/2},Q) f_b(A_- e^{-\yb/2},Q) \,,
\eeq
with 
\bea
\yb^- &=& {\rm max} \left(\ln A_-^2,  -\yb_{\rm cut}\right)\,, \nonumber\\
\yb^+ &=& {\rm min} \left(\ln \frac{1}{A_+^2},  \yb_{\rm cut}\right).
\eea
If we take the limit $k_T\to 0$ then Eq.~(\ref{difflum_mod}) reduces to the parton luminosity computed in the soft limit, Eq.~(\ref{difflum}).

So far we have discussed how to take into account the complete kinematics in the PDFs for the hardest emission. Clearly, the matrix elements will also differ from their eikonal approximations and the gap fraction ($f = 1 - \d\sigma^{\overline{\rm{gap}}}/\d\sigma^{\rm{born}}$) involves
\beq
\frac{\d \sigma^{\overline{{\rm{gap}}}}}{\d Q \d Y} = \int \d k_T \, \d y' \, \d \phi \; |\m_{2\to 3}(\dy,Q,k)|^2
\tilde{\l}(\dy,Q,k)~,
\eeq
where we have suppressed parton indices for clarity.
Since both the matrix elements and the parton luminosities depend on the momentum of the emitted parton, $k$, we lose the convenient kinematic factorization of Eq.~(\ref{gapsxsec}). Importantly, it is the shift in the argument of the PDFs that dominates, and so we shall evaluate the matrix elements in the eikonal limit. To further simplify matters we can also
restore the kinematic factorization by approximating the integral of the parton luminosity by its value at a particular phase space point. Specifically, we write 
\beq
\frac{\d \sigma^{\overline{{\rm{gap}}}}}{\d Q \d Y} \approx \int \d k_T \, \d y' \, \d \phi \; |\m_{2\to 3}(\dy,Q,k)|_{{\rm{soft}}}^2 
\left.\tilde{\l}(\dy,Q,k)\right|_{k_T=\sqrt{Q_0Q}, \,y'=\alpha\Delta y}\label{modification}
\eeq
and the value of $\bar{Q}$ is determined by its azimuthal average:
\beq
\int_0^{2 \pi} \frac{d \phi}{2 \pi} \bar{Q} = -\frac{k_T^2}{8 Q} + \ord\left( k_T^4 \right)\,.
\eeq
For the transverse momentum, we have chosen the geometric mean of the integration limits, i.e. ${k}_T=\sqrt{Q_0Q}$.
\def\ltsim{\mathrel{\rlap{\lower3pt\hbox{\hskip1pt$\sim$}}\raise2pt\hbox{$<$}}}
 The rapidity value is determined by requiring the approximate result on the right hand side of~(\ref{modification}) to be as close as possible to its exact value on the left hand side. We keep $\alpha$ fixed as we vary $Q$, but allow it to vary with $\Delta y$ and typically find $\frac{1}{4}\ltsim \alpha \ltsim \frac{1}{3}$. 

The $\ord(\alpha_s)$ modified gap fraction is plotted in Fig.~\ref{fig:mod} with solid lines. The plot shows that with a one parameter fit we can construct a modified resummation whose first-order expansion reproduces the FO result very accurately. 
We stress that this modification of the parton luminosity does not affect the formal leading logarithmic accuracy of our calculation. Rather it corresponds to a particular choice of important sub-leading terms that is motivated by energy-momentum conservation: it is very reassuring that such a procedure reproduces so well the exact leading order result.

Using this modified resummed result we are now ready to complete the matching to LO (the matching corrections are now very small), estimate the theoretical uncertainty and then compare to data.

\section{Matched results and comparison to data}
We define our modified resummed cross section (for zero gluons outside the gap), i.e. the replacement of Eq.~(\ref{gapsxsec}), as
\bea \label{gapsxsec_mod}
\frac{\d^2 \sigma^{\rm mod}}{\d Q \, \d Y} &=& \frac{\rho \pi \as^2}{2 Q S}  \sum_{a,b,c,d}  \frac{1}{1+\delta_{ab}}\frac{1}{1+\delta_{cd}} \Big \{|\m^{{\rm born}}_{abcd}|^2\mathcal{L}_{ab}(\dy,Q^2)
+\left(|\m^{(0)}_{abcd}|^2-|\m^{{\rm born}}_{abcd}|^2\right)  \nonumber\\ &&\times\left.\tilde{\mathcal{L}}_{ab}\left(\dy,Q,k\right)\right|_{k_T=\sqrt{Q_0Q}, \,y'=\alpha\Delta y} \Big\}.
\eea
We then define a resummed gap fraction by adding the FO calculation and the modified resummation together, subtracting the expansion of the resummed expression to $\ord(\alpha_s)$:
\beq \label{matchedgap}
f^{\rm gap}_{{\rm{matched}}} = f^{\rm gap}_{LO}+ f^{\rm gap}_{{\rm{mod}}}- f^{\rm gap}_{{\rm{mod}},\alpha_s}.
\eeq
We also estimate the effects of non-global logarithms by multiplying the above expression by the $K$-factor defined in Eq.~(\ref{ng_K}). 
The calculation we have performed matches together a LO computation with a leading logarithmic one and so we expect it to have a considerable theoretical uncertainty.
Because we are considering the gap fraction,  renormalisation and factorization scale variations do not give the dominant contribution to the uncertainty. Parton-distribution-function effects also largely cancel in the ratio. The dominant source of uncertainty comes from higher logarithmic orders in the resummation. In particular, a leading logarithmic resummation does not fix the argument of the logarithms we are resumming. As an estimate of our theoretical uncertainty we then rescale the argument of the function~$\xi$, Eq.~(\ref{rc}):
\beq
\xi(Q_0,Q) \longrightarrow \xi(\gamma Q_0,Q)\,,
\eeq
with $\gamma$ allowed to vary in a range of order unity.
Motivated by the fact that next-to-leading logarithmic corrections to leading soft logarithms are typically found to be negative, we consider variations in the upward direction by, quite arbitrarily, a factor of~2, i.e. $1<\gamma<2$.

\begin{figure} 
\begin{center}
\includegraphics[width=0.495\textwidth, clip]{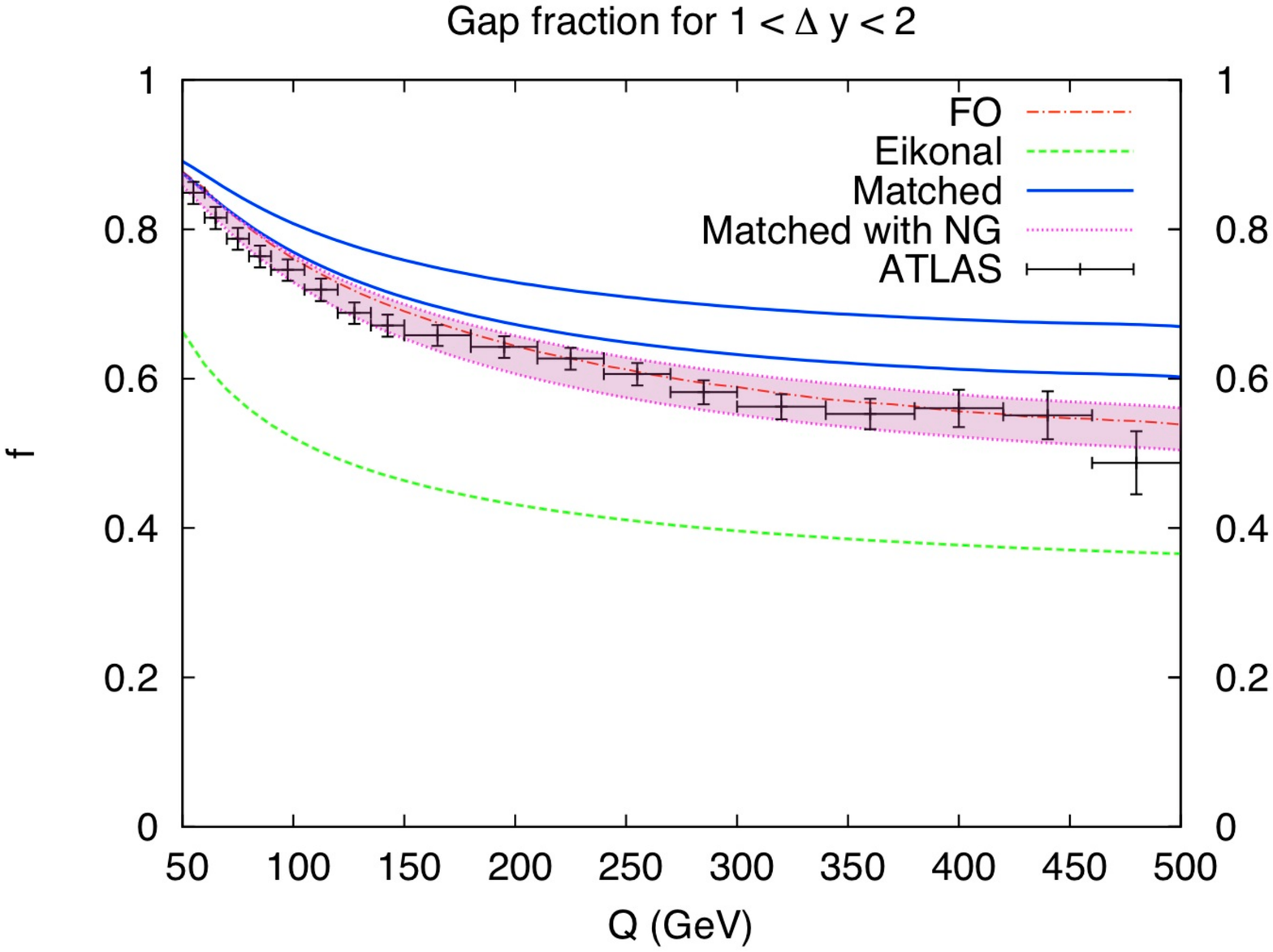}
\includegraphics[width=0.495\textwidth, clip]{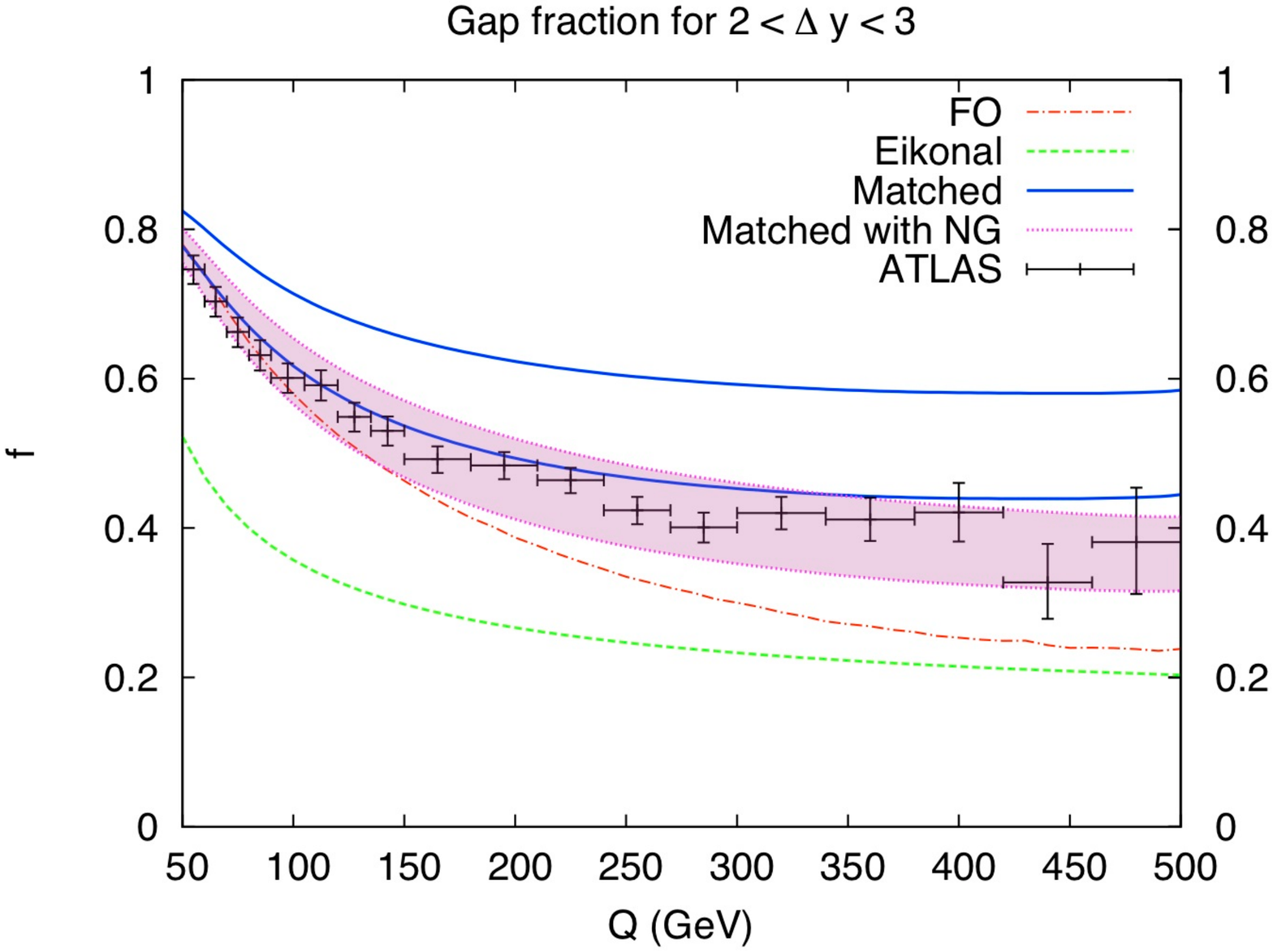}
\includegraphics[width=0.495\textwidth, clip]{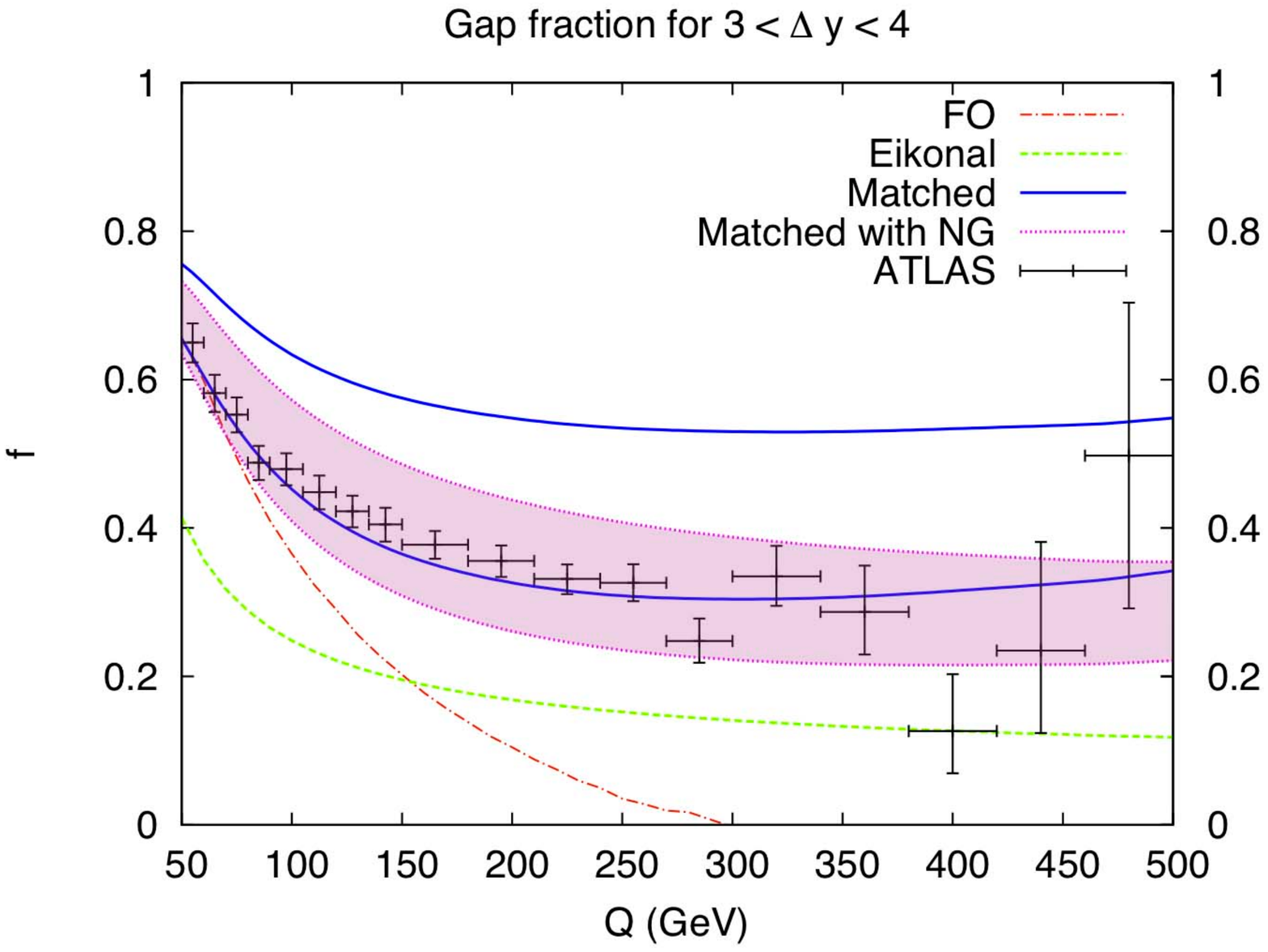}
\includegraphics[width=0.495\textwidth, clip]{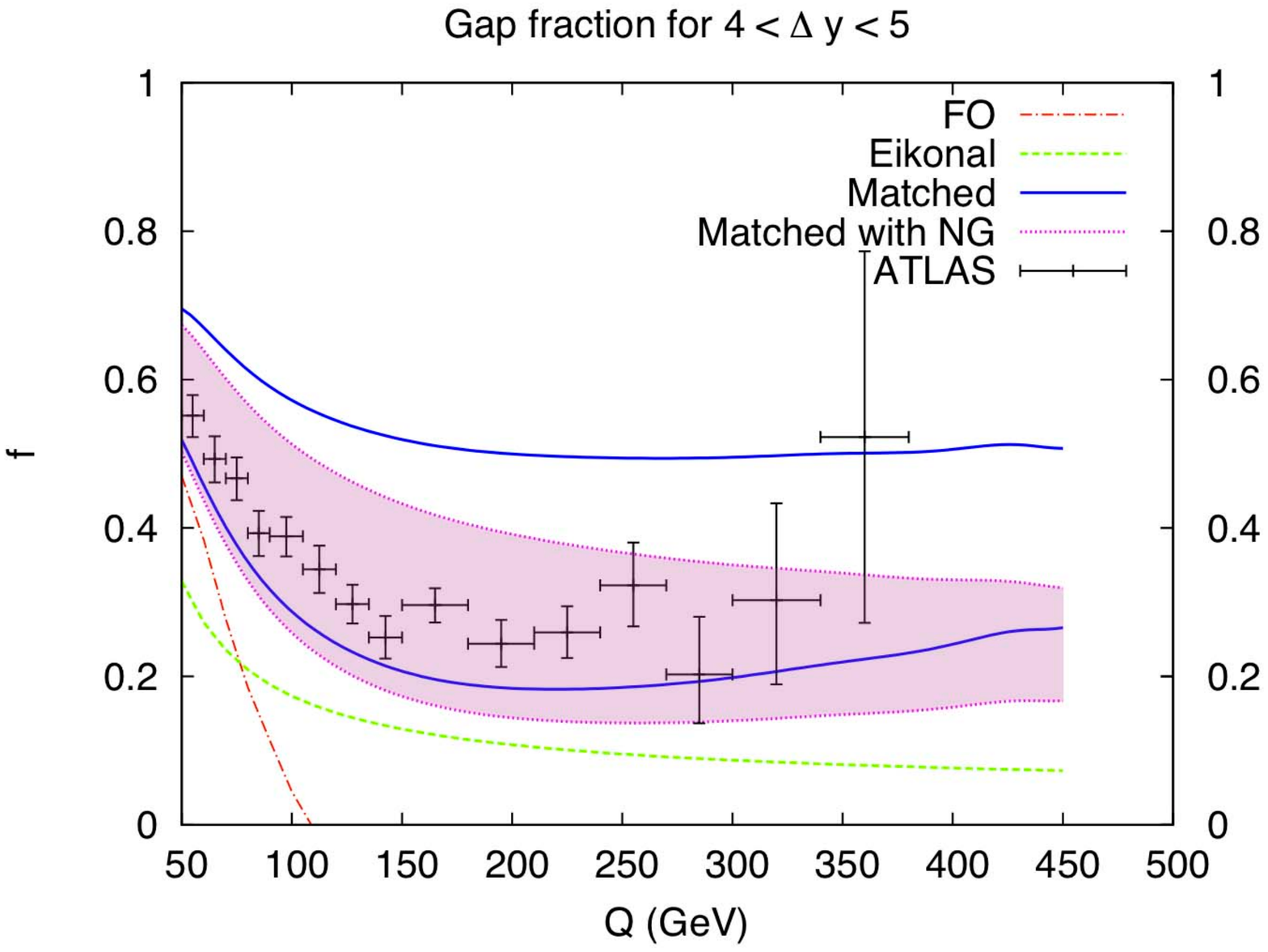}
\caption{The matched gap fraction as a function of the transverse momentum $Q$ in different rapidity bins. }\label{fig:match}
\end{center}
\end{figure}

\begin{figure} 
\begin{center}
\includegraphics[width=0.495\textwidth, clip]{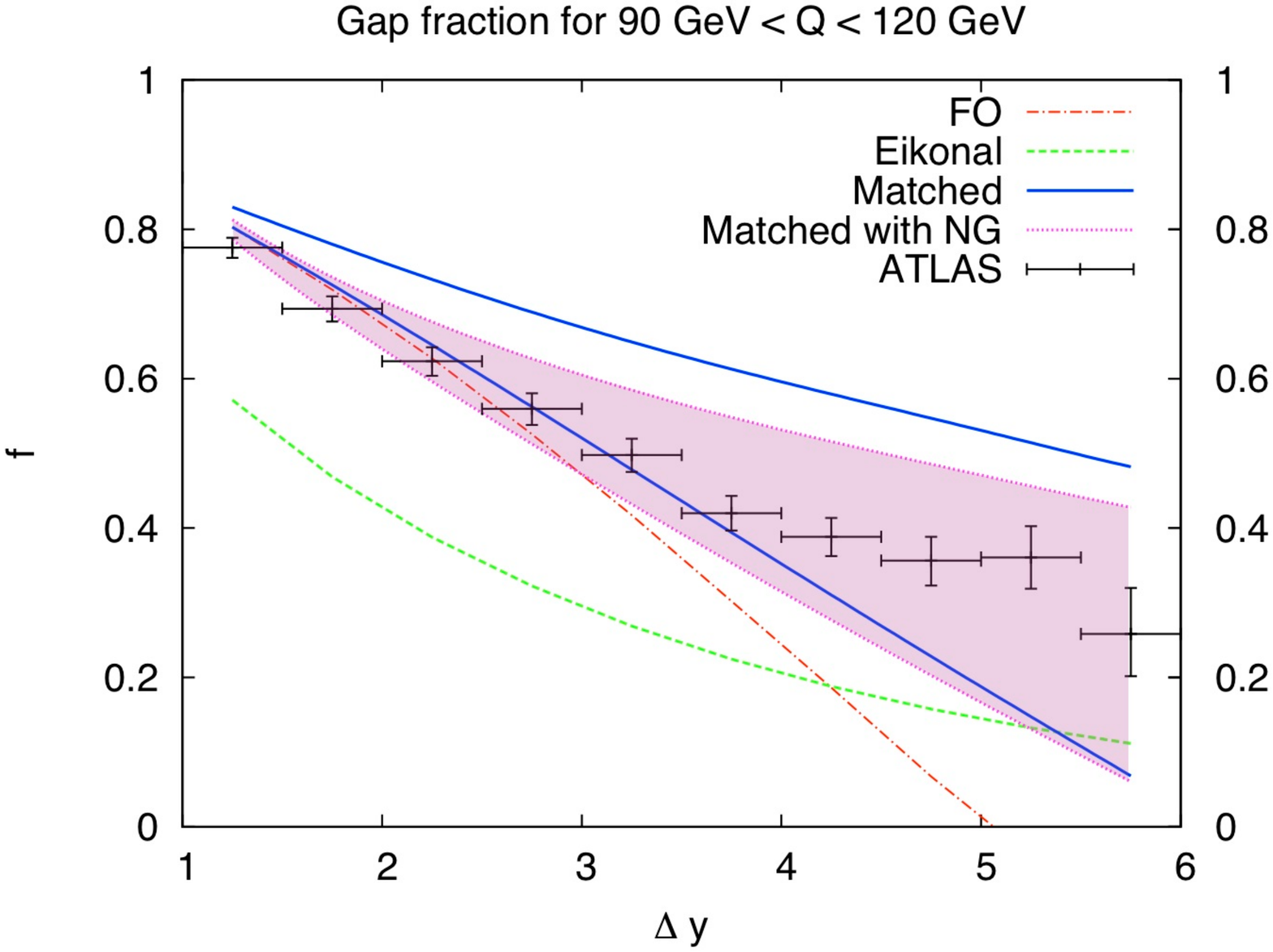}
\includegraphics[width=0.495\textwidth, clip]{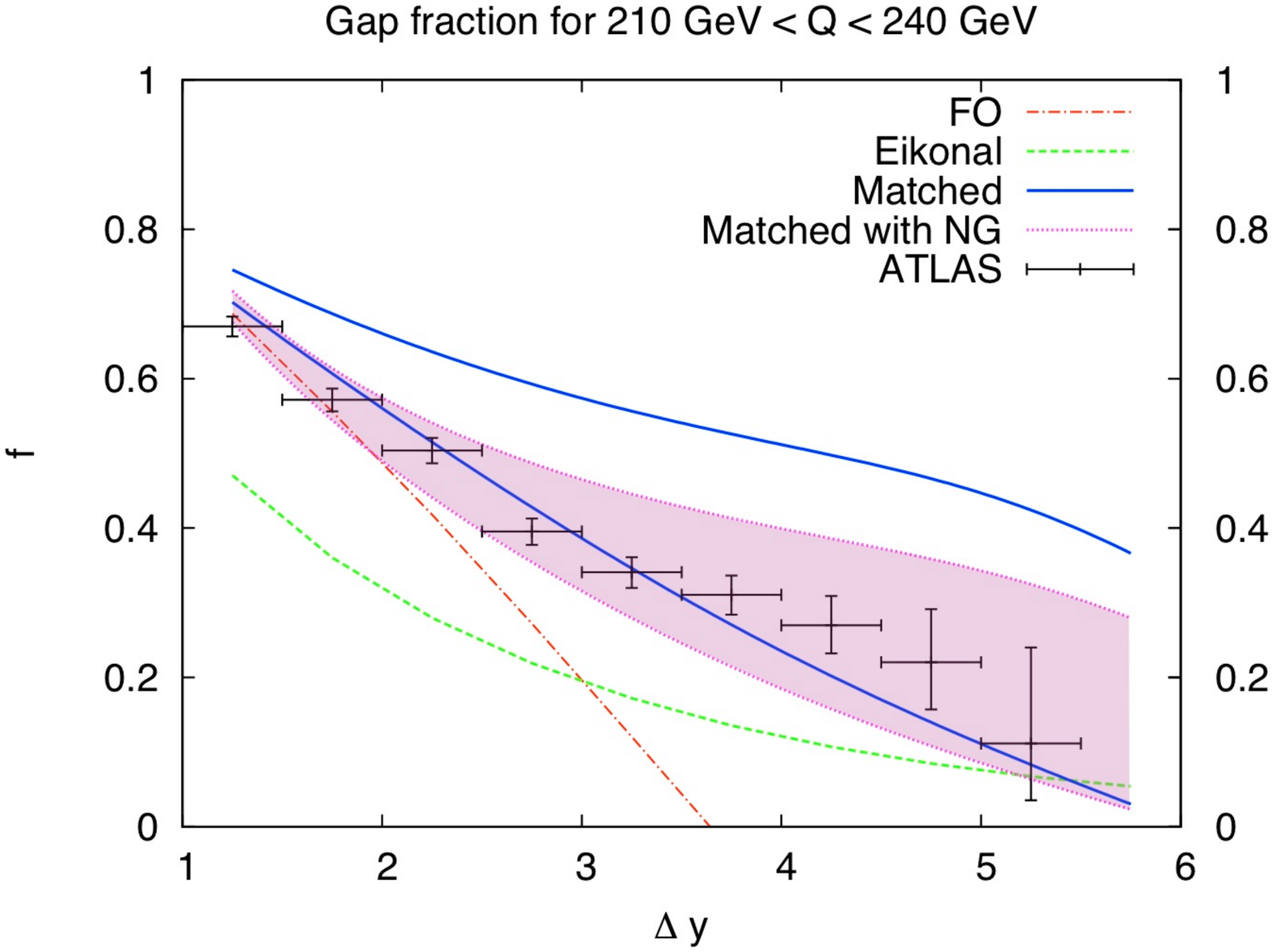}
\caption{The matched gap fraction as a function of rapidity separation $\Delta y$ in two different transverse momentum bins. }\label{fig:match_Dy}
\end{center}
\end{figure}

The plots in Fig.~\ref{fig:match} show the gap fraction as a function of the mean transverse momentum of the two leading jets $Q$ in four different rapidity bins, while the ones in Fig.~\ref{fig:match_Dy} are for the $\Delta y$ distribution, in two different $Q$ bins. We use the same cuts as the ATLAS analysis: all jets must have $p_T>20$~GeV, $|y|<4.4$ and the mean transverse momentum of the two highest transverse momentum jets must be $Q>50$~GeV. The dash-dotted red line represents the LO calculation, the dashed green line the resummed gap fraction in the eikonal limit, solid blue is the resummed and matched result Eq.~(\ref{matchedgap}), with the band obtained by varying $\gamma$, as explained above, and finally the magenta band corresponds to the resummed and matched gap fraction with the non-global effects included. The black crosses are the data points measured by the ATLAS collaboration~\cite{atlasveto} with the gap defined by the two highest $p_T$ jets (we have combined the statistical and systematic errors in quadrature). 

The FO calculation is clearly only sensible in the first rapidity bin and for $\Delta y > 2$ it decreases very rapidly as a function of $Q$ and eventually becomes negative. This unphysical behaviour is driven by a large logarithmic term $\sim \as \Delta y \ln \frac{Q}{Q_0}$ which needs to be resummed. The eikonal resummation restores the physical behaviour but, as we have previously discussed, completely ignores the issue of energy-momentum conservation and produces too small a gap fraction. Our matched curves, with the inclusion of non-global logarithms, does seem to capture most of the salient physics. However, our results are affected by large theoretical uncertainties due to the fact the calculation is accurate only at the leading logarithmic level. The extension of resummation for the gap cross section at the next-to-leading logarithmic accuracy will certainly reduce this uncertainty but it is not an easy task and it is not likely that it is going to be completed soon. Another way of reducing the uncertainty is to perform the matching at NLO, so that any dependency on the rescaling factor $\gamma$ is pushed one order higher in the strong coupling.
With the necessary NLO calculations available in \textsc{Nlojet}++~\cite{nlojet}, such a NLO matching is certainly feasible.

\section{Comparison to other approaches and conclusions}
In Ref.~\cite{atlasveto} comparisons are made between the data and the predictions of some of the different theoretical tools currently available.  Firstly we notice that gap fractions are defined there with respect to the dijet cross section at NLO, while we use the Born cross section. We have checked that, because of the definition of $Q$ as the mean transverse momentum, NLO corrections are small.
The best description of the data was found using POWHEG~\cite{powheg,2powheg,pbox}, interfaced with PYTHIA~\cite{pythia}. The results obtained using POWHEG interfaced with \textsc{Herwig}++~\cite{herwig} were found to undershoot the data. The difference between the two parton showers can be taken as indicative of the theoretical uncertainty due to the parton shower--NLO matching. The formal accuracy of the POWHEG calculations appearing in the ATLAS paper is not different to ours: Tree-level matrix elements are used and then matched to a parton shower, which is essentially a leading logarithmic resummation. However, the final predictions differ from ours because of the assumptions and approximations contained in the showering algorithm. Firstly, energy-momentum conservation is properly accounted for in every emission in a parton shower, not just the hardest as in our calculation. Also, the parton shower is limited to the large $N_c$ approximation, but it does include non-global logarithms beyond the ``one out-of-gap gluon approximation''.  
Another effect which is missing in the parton shower approach is Coulomb gluon exchange. As pointed out in~\cite{Forshaw:2009fz} these contributions are especially important in the large $Q/Q_0$ and large $\Delta y$ region: Coulomb gluons contribute to building up the colour-singlet exchange contribution, which eventually leads to a rise of the gap fraction at large enough $\Delta y$.  

The ATLAS collaboration also compared their data to theoretical predictions obtained with HEJ~\cite{HEJ}. That framework is based on the factorization of multi-gluon amplitudes in the high-energy regime.  As in the BFKL approach, $\as^n \Delta y^n $ terms are resummed, but energy-momentum conservation is enforced. Logarithms of $Q/Q_0$ are not systematically resummed, unless they come with a $\Delta y $ factor. We notice that the HEJ predictions are similar to ours for the global part, after accounting for energy-momentum conservation. This does not come as a surprise: Although the two approaches resum different terms, the leading contributions are of the form $\as^n \Delta y^n \ln^n \frac{Q}{Q_0}$ and are resummed in both approaches. HEJ describes emissions of out-of-gap gluons and, if interfaced with a parton shower, should be capable of capturing non-global logarithms as well ~\cite{hejPS}. We note that the HEJ framework does not at present include colour mixing via Coulomb gluons.

It seems clear that within the context of the overall accuracy of a leading log/leading order matching and the kinematic range of the current data, the impact of sub-leading $N_c$ and Coulomb gluon effects is not yet critical (except perhaps at the largest values of $\Delta y$ where both PYTHIA and \textsc{Herwig}++ undershoot the data). The same cannot be said about the constraints of energy-momentum conservation, which are clearly very important. The message is clear: the accuracy of the ATLAS data already demands better theoretical calculations.
\vspace{0.5 cm}

{\bf Acknowledgements} This work was supported by UK's STFC. We wish to thank Andrea Banfi, Mrinal Dasgupta and Andrew Pilkington for many useful discussions. MHS is supported in part by an IPPP Associateship.

\end{document}